\documentclass[11pt]{article}

\usepackage[
    colorlinks=true,
    linkcolor=black,    
    citecolor=black,    
    urlcolor=blue       
]{hyperref}

\usepackage{amsthm,amsmath,amssymb}
\usepackage{enumerate}

\usepackage[a4paper,margin=3 cm]{geometry}
\usepackage[numbers]{natbib}
\usepackage[normalem]{ulem}

\newtheorem{dfn}{Definition}[section]
\newtheorem{tw}[dfn]{Theorem}

\newtheorem{prop}[dfn]{Proposition}
\newtheorem{rem}[dfn]{Remark}

\newtheorem{lem}[dfn]{Lemma}

\newtheorem{cor}[dfn]{Corollary}

\numberwithin{equation}{section}

\newcommand*{\QQ}{\mathbb{Q}}

\DeclareMathOperator{\EE}{\mathbb{E}} 
\DeclareMathOperator{\PP}{\mathbb{P}} 

\global\long\def\dd#1{\textnormal{d}#1}

\title{An Explicit Distributional Representation for the SABR Model via Explosive Diffusions}

\author{Micha\l \ Barski  \\ \small  Faculty of Mathematics, Warsaw University, Poland\\
 \small{\it m.barski@mimuw.edu.pl} \bigskip \\
Bart\l omiej Polaczyk  \\ \small  Faculty of Mathematics, Warsaw University, Poland\\
 \small{\it b.polaczyk@mimuw.edu.pl} \bigskip \\
Maciej Wi\'sniewolski  \\ \small  Faculty of Mathematics, Warsaw University, Poland\\
 \small{\it wisniewolski@mimuw.edu.pl}}

\begin{document}

\maketitle

\begin{abstract}
We develop a new probabilistic representation of the asset price distribution in the widely used SABR model. Our approach reduces the original system of stochastic differential equations to an auxiliary one-dimensional diffusion with possible explosion, which is analyzed using techniques developed by Karatzas and Ruf \cite{KaratzasRuf}. We first derive explicit formulas for power transforms of the asset price and then invert them to obtain an exact representation of its distribution as an explicit functional of two independent Brownian motions. The resulting characterization leads to a Black--Scholes mixture representation for European call prices, providing both a probabilistic interpretation of the SABR model and a Rao--Blackwellized Monte Carlo method requiring simulation only of a one-dimensional diffusion. We further investigate the explosion time of the auxiliary process and characterize the associated survival probabilities, obtaining explicit formulas in the uncorrelated case. These results yield new analytical and computational insights into the probabilistic structure of the SABR model and its applications to option pricing and implied volatility analysis.
\end{abstract}

\noindent
\begin{quote}


\end{quote}

\section{Introduction}

In the seminal paper \cite{Hagan}, Hagan and co-authors introduced a stochastic volatility model known as SABR, an acronym derived from its key parameters: stochastic alpha, beta, and rho. The model describes the price dynamics of a financial asset by allowing its volatility to evolve stochastically. This is formalized through a system of two stochastic differential equations (SDEs): 
\begin{align}\label{rownanie wstep}
	dS_t = S_t^{\beta}Y_tdW_t, \quad dY_t = \alpha Y_tdZ_t, \quad S_0=s_0, Y_0=y_0,  \quad t\geq 0,
\end{align}
based on two $\rho$-correlated Wiener processes $W$ and $Z$ i.e. $\langle W,Z\rangle_t=\rho t$ with $\rho\in[-1,1]$.
Although the model's specification is relatively simple, the rigorous mathematical characterization of the solution to these SDEs has proven unexpectedly challenging. While the asset price distribution has been characterized explicitly in several special parameter regimes and analytical frameworks, a general probabilistic representation valid across broader parameter configurations has remained elusive. Despite this, SABR has become one of the most widely used models capable of capturing the empirically observed phenomenon known as the “volatility smile” in financial markets. This feature of the option market reflects the dependence of vanilla option prices on both the time to maturity and the strike prices, mediated through the volatility parameter. Mathematically, this relationship induces the geometrical structure: 
\begin{align*}
	(t,K) \mapsto \sigma(t,K), \quad t \in [0,T], \quad K\in [K_l,K_r],
\end{align*} 
called a volatility surface, which is essential in financial practice, as it serves as a key input for pricing large portfolios of derivative instruments handled by financial institutions.

The description of the SABR model in the sense of characterization of its distribution or, equivalently, call option prices is a field of active research with a rapidly increasing number of records. The discussion below is necessarily selective and focuses on results most closely related to the present work. There are several approaches using various techniques which provide an approximate solution of the problem or exact solution but for particular parameter regimes.  The original quantitative approach of Hagan et al. ~\cite{Hagan} to the model relied on derivation and approximate solving the forward Fokker-Planck equation for the transition density of the base asset by singular perturbation methods (cf. also~\cite{Hag1}). The resulting formula of Hagan for the Black-Scholes implied volatility is currently a  standard pricing tool for plain vanilla options, but also other approximation formulas of similar fashion were obtained in~\cite{Doust,Balland2013,Andreasen2013}, by means of short maturity expansion technique. It is well known that, in certain parameter regimes (notably for low strikes), the Hagan asymptotic formula may lose accuracy and can lead to violations of positivity of the implied density. The refinements based on a correction of the leading order term ~\cite{Obloj} or on computing the second order term ~\cite{Paulot} in the original expansion provide better  approximations in many regimes, but are not designed to fully capture the degeneracy of the model near zero. Exact formula for the density function of the absolute continuous part of the distribution of $S$ on $(0,+\infty)$ was obtained in ~\cite{Ant} in the case $\rho=0$. The approach was extended for the lognormal SABR model with negative correlation, i.e. when $\beta=1$ and $\rho<0$, in ~\cite{fordezhang} and yields some large strike  asymptotic expansion for the implied volatility. The zero correlation case has several other characterizations in the literature. In ~\cite{islah} an explicit option pricing formula is derived in terms of multidimensional integrals. Another approach exploits the connection of the SABR model with the Constant Elasticity of Variance (CEV) model 
\begin{gather}\label{CEV equation}
dS_t=S^\beta_t dW_t, \quad S_0=s_0, \quad t\geq 0.
\end{gather}
The transformation 
\begin{gather}\label{CEV transformation}
X_t:=\frac{S_t^{2(1-\beta)}}{(1-\beta)^2},
\end{gather}
yields a squared Bessel process and enables computing prices of call options in the CEV model. As SABR is a CEV model enriched by stochastic volatility, the method can be extended for option pricing in SABR but requires also the $\rho=0$ condition, see ~\cite{antonov2014exact}. The uncorrelated case was considered also in ~\cite{antonov2012advanced} 
where option prices are represented in terms of two dimensional integrals of elementary functions from which some approximations for the 
$\rho\neq 0$ case are also derived. In ~\cite{Doust}, ~\cite{Gul}, ~\cite{gulisashvili2018mass} the singular part of the distribution of $S$ is studied with emphasis on  refinements of the Hagan formula close to zero for particular parameter setup. Explicit formulas for the probability of hitting zero were obtained in the special case with $\rho=\beta=0$ in terms of the Bessel functions in ~\cite{Gul}. The irregularity of the SABR model at zero affects also numerical pricing methods. Since the transformation $x\longrightarrow x^\beta$ is not Lipschitz continuous at zero for $\beta\in(0,1)$, the application of the Euler scheme for $S$ in \eqref{rownanie wstep} is problematic, see ~\cite{chen2012low}, so is the option pricing method based on Monte Carlo simulations. Similarly, the lack of regularity affects also finite difference methods and finite element methods used for option pricing and give rise to deep modifications, like in ~\cite{horvath18sifin}, where the pricing Kolmogorov equation is discretized via non-symmetric Dirichlet forms and the setup is based on certain Sobolev spaces with singular weights. In the geometric view on the SABR model, which became popular in recent time ~\cite{hagan2015probability}, ~\cite{henry2006unifying}, 
 ~\cite{Gul} the process $(S,Y)$ in \eqref{sabr rownanie} is transformed by a change of time to a Brownian motion on a Riemannian manifold. This approach is suitable for the cases where $\beta=0$ or $\rho=0$ and allows to rederive the expansion formula of Hagan. Since for $\beta\in(0,1)$ the description brakes down at zero, the approach required modifications. Horvath and collaborators \cite{Do} embedded the SABR model into a Hilbert space setting, solving the Kolmogorov equation using symmetric Dirichlet forms. They established the right functional-analytic setting for pricing numerics: they derived Feller-type properties and strong continuity results for the SABR semigroup, and identified suitable Banach/Hilbert spaces where these properties hold - precisely the kind of assumptions numerical schemes rely on.  They also complemented the “heat-kernel geometric”  viewpoint with a more general semigroup framework, and, in certain parameter configurations, they recovered an appropriate geometry via symmetric Dirichlet forms instead.
 A comprehensive overview of both theoretical and practical developments of the SABR model is available in Antonov’s monograph \cite{Ant}.

In this work, we derive an explicit probabilistic representation of the SABR marginal distribution in the general parameter setting. The key role in our construction is played by the ratio process $\theta$ defined in \eqref{eq:theta-def}, which turns out to be a one-dimensional diffusion with possible explosion. An unconventional change of probability measure, whose density depends explicitly on the Mellin parameter, allows us to express the Mellin transform of the asset price in terms of this process. Using the framework of Karatzas and Ruf \cite{KaratzasRuf}, we derive an explicit representation of the Mellin transform and subsequently invert it by separating the parameter dependence in the drift of the ratio process. This yields a representation of the SABR marginal distribution as an explicit functional of a one-dimensional auxiliary diffusion and an independent Brownian motion. To the best of our knowledge, this is the first probabilistic representation of the SABR distribution of this type. This representation admits also a direct financial interpretation. In particular, we show that European call prices can be represented exactly as expectations of Black--Scholes prices with random effective spot and random total variance generated by the auxiliary one-dimensional diffusion. This  Black--Scholes mixture representation naturally leads to a Rao--Blackwellized Monte Carlo algorithm, in which only the auxiliary one-dimensional diffusion needs to be simulated while the remaining Gaussian randomness is integrated out analytically. Consequently, the resulting estimator is unbiased and enjoys a rigorous variance reduction property. The same representation also provides a probabilistic characterization of the associated Black--Scholes implied volatility and yields an exact benchmark complementary to the classical asymptotic approximation of Hagan et al.

We additionally investigate the explosion time of the auxiliary diffusion arising under the change of measure. This leads to a characterization of survival and non-explosion probabilities through backward Kolmogorov equations and yields explicit formulas in the uncorrelated case.

The principal contributions of the paper may be summarized as follows.

\begin{itemize}

\item We derive an explicit probabilistic representation of the SABR marginal distribution as a functional of a one-dimensional auxiliary diffusion and Brownian motion, thereby reducing the analysis of the original two-dimensional stochastic volatility model to a one-dimensional diffusion problem.

\item We show that the representation induces an exact Black--Scholes mixture formula for European call prices, providing a direct financial interpretation of the probabilistic construction and yielding a natural Rao--Blackwellized Monte Carlo estimator together with a probabilistic characterization of the implied volatility.

\item We analyze the explosion time of the associated auxiliary
diffusion and characterize survival and non-explosion probabilities,
obtaining explicit formulas in the zero-correlation case.

\end{itemize}

During the revision of the present manuscript, we became aware of the
recent work of Pirjol and Zhu \cite{pirjol2025vix}, where the same
one-dimensional reduction of the SABR dynamics through the ratio process \eqref{eq:theta-def}
is employed in the study of VIX options and explosion properties of the
associated diffusion. However, the main idea of the present manuscript was originally completed
 in May 2024. While the one-dimensional reduction is
common to both approaches, our objectives and results are substantially
different.

The paper is organized as follows. Section~\ref{section Preliminaries} introduces the model and the notation. Section \ref{section Main results} develops the probabilistic
representation of the SABR distribution together with its consequences
for option pricing, including an exact Black--Scholes mixture
representation and a Rao--Blackwellized Monte Carlo interpretation.
 All proofs are
collected in Section \ref{Section Proofs}.

\section{Preliminaries}\label{section Preliminaries}
Let $(\Omega,\mathcal{F},\mathbb{P})$ be a probability space endowed with a filtration $(\mathcal{F}_t)_{t\geq 0}$ satisfying usual conditions and $(V,Z)$ be a two-dimensional Brownian motion with independent coordinates adapted to $(\mathcal{F}_t)$. Then the pair $(W,Z)$ with 
\begin{gather}\label{Wiener korelacja}
	W_t:= \rho Z_t + \sqrt{1 - \rho^2}V_t, \quad t\geq 0,
\end{gather}
where $\rho\in[-1,1]$, defines correlated Brownian motions. We consider the following system of SDEs called the SABR model:
\begin{align}\label{sabr rownanie}
    \begin{cases}
	    dS_t = S_t^{\beta}Y_tdW_t,\\
	    dY_t = \alpha Y_tdZ_t, 
    \end{cases}
	t\geq 0, 
\end{align}
with $Y_0=y_0>0, S_0=s_0>0$, where $\beta \in [0,1)$ and $\alpha > 0$ are fixed parameters. The process $Y$ is a stochastic exponential 
of $Z$ and plays the role of a random factor making the asset price $S$ of stochastic volatility type. The existence of $S$ satisfying \eqref{sabr rownanie} follows from standard arguments based on the change of time
\begin{gather}\label{change of time}
 \tau_t := \inf\{u\geq 0: \int_0^uY_s^2ds > t\}.
\end{gather}
\noindent
Let
\begin{gather}\label{hitting time}
\sigma:=\inf\{t\geq 0: S_t=0\},
\end{gather}
denote the first hitting time of zero by $S$.
Introducing the transformation
\[
X_t=\frac1{(1-\beta)^2}(S_{t\wedge\sigma})^{2(1-\beta)},
\]
It\^o's formula yields
\[
dX_t
=
2\sqrt{X_t}Y_t
\mathbf1_{[0,\sigma)}(t)dW_t
+
cY_t^2
\mathbf1_{[0,\sigma)}(t)dt,
\]
where
\[
c=\frac{1-2\beta}{1-\beta}.
\]
Consequently, after the time change
\[
u\mapsto\int_0^uY_s^2ds,
\]
the process $X$ becomes a squared Bessel process of dimension
$c\le1$, which therefore reaches zero almost surely.
Throughout the paper we assume that $S$ (equivalently $X$) is absorbed after
hitting zero.

\section{Main results}\label{section Main results}

The main observation underlying the present work is that the SABR model
admits a reduction to a one-dimensional diffusion through an appropriate
transformation. More precisely, our analysis is based on the {\it ratio
process}
\begin{equation}\label{eq:theta-def}
    \theta_t:=(1-\beta)\frac{Y_t}{S_t^{1-\beta}},
    \qquad t\ge0,
\end{equation}
which starts from the positive value
\[
    \theta_0=(1-\beta)\frac{y_0}{s_0^{1-\beta}}.
\]
Since $Y$ is a stochastic exponential and therefore remains strictly
positive, the process $\theta$ stays positive until the first time that
$S$ reaches zero. Consequently, the first hitting time of zero by $S$
coincides with the explosion time of $\theta$:
\[
    \sigma
    =
    \inf\{t\ge0:\theta_t=+\infty\},
\]
where $\sigma$ is defined in \eqref{hitting time}. Thus, after identifying
the dynamics of $\theta$, the study of the original two-dimensional SABR
model may be replaced by the analysis of a one-dimensional explosive
diffusion.

The remainder of the paper is devoted to exploiting the reduction
established in Theorem~\ref{thm:main}.
First, we derive functional representations of moments and
distributions. Next, we obtain probabilistic representations in terms
of Brownian motion and discuss applications to option pricing through a
Black--Scholes mixture representation.

The following theorem constitutes the main structural result of the paper.
It shows that every positive moment of the SABR price can be represented as
a negative moment of the auxiliary diffusion $\theta$ under an equivalent
change of measure. This reduction serves as the basis for all subsequent
functional representations developed in the paper.

\begin{tw}[One-dimensional reduction of SABR]
\label{thm:main}
Let us assume that zero is an absorbing state for $S$. Then, for any $p>0$, $\EE S_t^{(1-\beta)p} <+\infty$ and there exist a diffusion $\theta$ and an equivalent measure $\QQ=\QQ^{(p)}$, such that 
for any $t\geq0$, 
\begin{align}\label{rownanie na momenty}
	\EE
    S_t^{(1-\beta)p} 
    = 
    (1-\beta)^p
    e^{\frac{\alpha^2p(p-1)t}{2}}
    \EE_\QQ \theta_t^{-p}.
\end{align}
Moreover, $\mathbb{Q}$ is given by the density 
\begin{align}\label{Q densityyy}
    \frac{\dd \QQ}{\dd \PP}|_{\mathcal{F}_t} 
    = 
    e^{p\alpha Z_t - \frac{p^2\alpha^2}{2}t},
    \qquad t\ge 0.
\end{align}
and $\theta$ satisfies the following SDE
\begin{gather}\label{rownanie na teta}
	d\theta_t = \sigma(\theta_t)dB_t + \mu_p(\theta_t) dt,
	\quad
    t\ge 0,
\end{gather}
with
\begin{equation}\label{definicja sigma i mu}
\begin{aligned}
	\sigma(x) &:= \sqrt{x^4 - 2\alpha\rho x^3 +\alpha^2 x^2},\\ 
\mu_p(x) &:= \alpha^2 px - \alpha \rho(p+1) x^2 +\frac{2-\beta}{2-2\beta} x^3,\quad x\geq 0,
\end{aligned}
\end{equation}
where $B$ is a Brownian motion 
 under $\mathbb{Q}$.
\end{tw}

Identity \eqref{rownanie na momenty} replaces the computation of
moments of the two-dimensional SABR process by the computation of
negative moments of the one-dimensional diffusion $\theta$.
This reduction is the cornerstone of all subsequent distributional and
functional representations obtained in the paper. It is worth
emphasizing that the dynamics \eqref{rownanie na teta} of $\theta$ depend on
the order $p$ of the moment under consideration through the drift
coefficient $\mu_p$. In accordance with our convention, the solution of
\eqref{rownanie na teta} is understood up to its explosion time, at which
$\theta$ is absorbed at $+\infty$.

\begin{rem}\label{rem lognormal SABR}[Lognormal SABR]
Theorem~\ref{thm:main} and all subsequent results concern the case
$\beta\in[0,1)$. In the lognormal SABR model ($\beta=1$), analogous
arguments yield the identity
\[
    \EE (\ln S_t)^p
    =
    e^{\frac{p(p-1)t}{2}}
    \EE_{\QQ}\theta_t^p,
\]
where
\[
    \theta_t:=\frac{\ln S_t}{Y_t},
\]
and the process $\theta$ satisfies the stochastic differential equation
\begin{equation}\label{eq:theta-when-beta=1}
    d\theta_t
    =
    \sqrt{1-2\rho\theta_t+\theta_t^2}\,dB_t
    +
    \left(
        (1-p)(\theta_t-\rho)
        -\frac12
        e^{B_t+\frac{2p-1}{2}t}
    \right)dt.
\end{equation}
The corresponding characterization of $\theta$ is presented in
Section~\ref{Section Proofs}.
\end{rem}

\subsection{Functional representations via one-dimensional reduction}\label{subsection Functional probabilistic representation of the SABR model}

Building on the one-dimensional reduction established in
Theorem~\ref{thm:main}, we now derive functional representations for the
moments and the marginal distribution of the SABR price process. The main
idea is to remove the drift of the auxiliary diffusion $\theta$ by means of
a generalized Girsanov theorem for explosive diffusions
\citep[Theorem~3.2]{KaratzasRuf}, and then to apply a Lamperti-type
transformation reducing the problem to functionals of Brownian motion.

We first consider the driftless counterpart of
\eqref{rownanie na teta}--\eqref{definicja sigma i mu},
\begin{gather}\label{theta trifftless}
    d\theta_t^o=\sigma(\theta_t^o)dB_t^o,
    \qquad t>0,
\end{gather}
with the same initial condition as $\theta$, namely
\[
    \theta_0^o=\theta_0.
\]
The applicability of the explosive diffusion approach of
\cite{KaratzasRuf} requires the standing assumption
\begin{gather}\label{KarRuf standing assumption}
\int_K
\frac{1+|\mu_p(x)|}{\sigma^2(x)}\,dx<+\infty,
\qquad
\text{for every compact set }K\subset I:=(0,+\infty).
\end{gather}
This condition is satisfied for all $\rho\in[-1,1)$. For $\rho=1$, the
coefficient $\sigma$ has an additional zero at $x=\alpha$, and therefore
$1/\sigma^2$ is not locally integrable around this point.

Consequently, Theorem~3.2 of \cite{KaratzasRuf} can be applied on the state
space $I=(0,+\infty)$. In particular, there exists a weak solution
$(\theta^o,B^o)$ to \eqref{theta trifftless}, defined up to its explosion
time $\sigma^o$, understood as the exit time from $I$. Since $\theta^o$
cannot explode to $+\infty$ (see \cite[p.~332]{KS}), we have
\begin{gather}\label{driftless theta hits zero}
    \sigma^o=\inf\{t>0:\theta_t^o\le0\},
\end{gather}
that is, $\sigma^o$ is the first hitting time of zero by $\theta^o$.

The generalized Girsanov theorem yields a representation of the moments of
$S$ as functionals of $\theta^o$, see \eqref{firstexplo}. A second
transformation then reveals a Brownian structure. More precisely, we set
\[
    U_t:=\varphi(\theta_t),
\]
where $\varphi$ is chosen so that $U$ has unit diffusion coefficient:
\begin{gather}\label{U funkcja w opisie}
    U_t=u_0+B_t+\int_0^t b_p(U_s)\,ds,
    \qquad t\ge0.
\end{gather}
The driftless counterpart of $U$ is then a shifted Brownian motion. Applying
the generalized Girsanov theorem once more gives the Brownian functional
representation of the moments of $S$, see \eqref{wzor w T1}, as well as the
distribution of the first hitting time of zero by $S$, see
\eqref{rozklad dijscia do zera przez S}.

Finally, since the hitting time of zero by $S$ coincides with the explosion
time of $\theta$, Feller's test for explosions can be applied to the
one-dimensional diffusion \eqref{rownanie na teta}. This yields positive
probability of absorption at zero in finite time, see
\eqref{S reaches zero in finite time}.

To implement the Lamperti-type transformation explicitly, we now compute the
change of variable function $\varphi$ and the induced drift coefficient
$b_p$ in terms of the SABR parameters. Requiring the transformed process
\[
    U_t=\varphi(\theta_t)
\]
to have unit diffusion coefficient forces $\varphi$ to satisfy
\begin{gather}\label{varphi rownanie}
    \varphi'(x)\sigma(x)=1,
    \qquad
    \varphi(x_0)=\varphi_0,
    \qquad x>0.
\end{gather}
The solution of \eqref{varphi rownanie} is a strictly increasing $C^2$-class function on $(0,+\infty)$ given by 
\begin{align}\label{varphi wzooorr}
\varphi(x)&=\frac{1}{2}\ln \left(\frac{\sqrt{x^2-2\rho\alpha x+\alpha^2}-(\alpha-\rho x)}{\sqrt{x^2-2\rho\alpha x+\alpha^2}+(\alpha-\rho x)}\right), \ x>0, \quad \text{if} \quad \rho\in[-1,1).
\end{align}
For $\rho\in(-1,1)$ the limits of $\varphi$ at the boundary are
\begin{gather}\label{granice varphi}
\lim_{x\rightarrow 0^{+}}\varphi(x)=-\infty, \qquad  \lim_{x\rightarrow +\infty}\varphi(x)=\frac{1}{2}\ln \left(\frac{1+\rho}{1-\rho}\right)=:\gamma.
\end{gather}
For the formulation of the results announced above  we need also the inverse function $\varphi^{-1}:(-\infty,\gamma)\rightarrow (0,+\infty)$ which is given by
\begin{align}
\varphi^{-1}(x)&=\alpha\cdot\frac{4\rho e^{2x}+2\sqrt{1-\rho^2}e^x(1-e^{2x})}{(\rho^2-1)(e^{4x}+1)+2(\rho^2+1)e^{2x}},  \qquad  x\in(-\infty,\gamma),
\end{align}
and two auxiliary functions $h_p$ and $b_p$, with $p>0$, defined by  
\begin{gather}\label{wzor na h}
    h_p(x):=
    \frac12\varphi^{\prime\prime}(x)\sigma^2(x)
    +
    \frac{\mu_p(x)}{\sigma(x)},
    \qquad x>0.
\end{gather}
and 
\begin{align}\label{wzor na b_p}
b_p(x):=(h_p\circ \varphi^{-1})(x) \ 
, \quad x\in(-\infty,\gamma).
\end{align}
Actually, the inverse may be expressed
in the simpler form.

\begin{prop}[Explicit form of the transformed drift]
\label{prop:explicit-bp}
Let $\rho\in(-1,1)$ and denote
\[
\delta:=\sqrt{1-\rho^{2}},
\qquad
\gamma
:=
\frac12
\log\!\left(
\frac{1+\rho}{1-\rho}
\right).
\]
Then the inverse Lamperti transform admits the representation
\[
\varphi^{-1}(u)
=
\frac{\alpha}
{\rho-\delta\sinh u},
\qquad
u<\gamma.
\]
Moreover, the drift coefficient
\[
b_p=h_p\circ\varphi^{-1}
\]
can be written explicitly as
\[
b_p(u)
=
\alpha
\left[
\frac{\beta}
{2(1-\beta)\,
\delta\cosh u\,
(\rho-\delta\sinh u)}
-
\left(
p-\frac12
\right)
\tanh u
\right],
\qquad
u<\gamma.
\]
In particular, for the normal SABR model
($\beta=0$),
\[
b_p(u)
=
-\alpha
\left(
p-\frac12
\right)
\tanh u,
\qquad
u<\gamma,
\]
and therefore
\[
|b_p(u)|
\le
\alpha
\left|
p-\frac12
\right|,
\qquad
u<\gamma.
\]
\end{prop}

In particular, the drift coefficient appearing in the Brownian representation
of Theorem~\ref{T1} is available in closed form and can be evaluated
without solving any auxiliary differential equation.

Below expectations under the measure $\mathbb{Q}=\mathbb{Q}^{(p)}$ defined by \eqref{Q densityyy}
and under the measure $\mathbb{Q}^{o}$ associated with the weak solution of \eqref{theta trifftless}
are denoted by $\mathbb{E}^{(p)}[\cdot]$ and $\mathbb{E}^{o}[\cdot]$, respectively.
\vskip1ex

\begin{tw}[Brownian functional representations for the SABR model]
\label{T1}
Let $\rho\in[-1,1)$ and let $(S_t)$ be given by \eqref{sabr rownanie}.
Then:
\begin{enumerate}
\item{} For any $p>0$ and $t>0$,
\begin{align}\label{firstexplo}
\mathbb{E}\left(\frac{S_t^{(1-\beta)p}}{(1-\beta)^p}\right)
&=
y_0^p e^{\frac{\alpha^2p(p-1)t}{2}}
\mathbb{E}^{o}\left[
e^{\int_0^t\nu_p(\theta_s^o)dB_s^o
-\frac12\int_0^t\nu_p^2(\theta_s^o)ds}
(\theta_t^o)^{-p}
\mathbf{1}_{\{t<\sigma^o\}}
\right],
\end{align}
where $(\theta^o)$ is given by \eqref{theta trifftless}, $\sigma^o$ by \eqref{driftless theta hits zero} and $\nu_p(x):=\mu_p(x)/\sigma(x)$, $x>0$. Moreover, the process $(\theta_t^o)$ admits the representation 
\begin{align}\label{Lampertiii}
    \theta^o_{\int_0^t
    \frac{1}{\sigma_0^2\left(\theta_0e^{W_s-\frac{s}{2}}\right)}\,ds}
    =
    \theta_0e^{W_t-\frac{t}{2}},
    \qquad t<\sigma^o,
\end{align}
where 
\[
    \sigma_0(x):=\sqrt{x^2-2\alpha\rho x+\alpha^2},
    \qquad x>0.
\]
 and $W$ is a standard Brownian motion under $\mathbb{Q}^{o}$. 
\item{} If $\rho \in (-1,1)$ and $\gamma$ is defined by \eqref{granice varphi},  then for any $p>0$ and $t>0$,
\begin{align}\label{wzor w T1}\nonumber
\mathbb{E}\left(\frac{S_t^{(1-\beta)p}}{(1-\beta)^p}\right)
&=
y_0^p e^{\frac{\alpha^2p(p-1)t}{2}}
\mathbb{E}\Bigg[
e^{\int_0^t b_p(u_0+W_s)dW_s
-\frac12\int_0^t b_p^2(u_0+W_s)ds}
\\
&\qquad\qquad\qquad \times
\mathbf{1}_{\{t<\sigma^W_{\gamma-u_0}\}}
\frac{1}{(\varphi^{-1}(u_0+W_t))^p}
\Bigg],
\end{align}
where $u_0:=\varphi((1-\beta)y_0/s_0^{1-\beta})$, $W$ is a standard Brownian motion under $\mathbb{P}$, $\sigma^W_{\gamma-u_0} = \inf\{t\geq 0: W_t = \gamma-u_0\}$.
Furthermore, under the measure $\mathbb{Q}^{(p)}$, the distribution of the first hitting time
of $\{0\}$ by $S$ is given by
\begin{align}\label{rozklad dijscia do zera przez S}
\mathbb{Q}^{(p)}(\sigma > t)= \mathbb{E}\left[e^{\int_{0}^{t}b_p(u_0+W_s)dW_s-\frac{1}{2}\int_{0}^{t}b_p^2(u_0+W_s)ds}\cdot  \mathbf{1}_{\{t<\sigma^{W}_{\gamma-u_0}\}}\right], \quad t\geq 0.
\end{align}
\item{} Finally, for every $\rho\in[-1,1)$ there exists $p>0$ such that
\begin{gather}\label{S reaches zero in finite time under Q}
    \mathbb{Q}^{(p)}(\sigma<+\infty)=1.
\end{gather}
Consequently,
\begin{gather}\label{S reaches zero in finite time}
    \mathbb{P}(\sigma<+\infty)>0.
\end{gather}
\end{enumerate}
\end{tw}

\begin{rem}[Kolmogorov characterization of the survival probability]
The survival probability
\[
v(t,u)
:=
\mathbb Q_u^{(p)}(\sigma>t),
\qquad
u<\gamma,
\]
may equivalently be characterized as the unique bounded solution of the
backward Kolmogorov equation
\[
\partial_t v
=
\frac12\partial_{uu}v
+
b_p(u)\partial_u v,
\qquad
t>0,\quad u<\gamma,
\]
subject to the initial and boundary conditions
\[
v(0,u)=1,
\qquad
v(t,\gamma)=0.
\]

Similarly, the non-explosion probability
\[
w(u)
:=
\mathbb Q_u^{(p)}(\sigma=\infty)
\]
satisfies the harmonic equation
\[
\frac12w''(u)+b_p(u)w'(u)=0,
\qquad
u<\gamma,
\]
with boundary condition
\[
w(\gamma)=0.
\]
In the particular case $\rho=0$, the explicit expression for $b_p$
derived in Proposition \ref{prop:explicit-bp} reduces these equations to
ordinary differential equations that can be solved in closed form.
\end{rem}

\begin{cor}[One-dimensional PDE characterization of SABR moments]
\label{cor:one-dimensional-PDE}
Let $\rho\in(-1,1)$ and $p>0$. Set
\[
    u_0
    :=
    \varphi\left(
        (1-\beta)\frac{y_0}{s_0^{1-\beta}}
    \right),
    \qquad
    \gamma
    :=
    \frac12
    \log\left(\frac{1+\rho}{1-\rho}\right),
\]
and define
\[
    g_p(u)
    :=
    \frac{1}{(\varphi^{-1}(u))^p},
    \qquad u<\gamma.
\]
Let $U$ be the diffusion
\[
    dU_t
    =
    dB_t+b_p(U_t)\,dt,
    \qquad U_0=u<\gamma,
\]
killed upon hitting $\gamma$, and denote
\[
    \sigma_\gamma^U
    :=
    \inf\{t\ge0:U_t=\gamma\}.
\]
Define
\[
    v_p(t,u)
    :=
    \mathbb E_u
    \left[
        g_p(U_t)
        \mathbf 1_{\{t<\sigma_\gamma^U\}}
    \right].
\]
Then $v_p$ solves the killed Cauchy problem
\begin{equation}\label{eq:pde-moment-vp}
\begin{cases}
    \partial_t v_p(t,u)
    =
    \frac12\partial_{uu}v_p(t,u)
    +
    b_p(u)\partial_u v_p(t,u),
    & t>0,\ u<\gamma,\\[0.6ex]
    v_p(0,u)=g_p(u),
    & u<\gamma,\\[0.6ex]
    v_p(t,\gamma)=0,
    & t\ge0.
\end{cases}
\end{equation}
Moreover,
\begin{equation}\label{eq:moment-pde-representation}
    \mathbb E S_t^{(1-\beta)p}
    =
    (1-\beta)^p
    e^{\frac{\alpha^2p(p-1)t}{2}}
    v_p(t,u_0).
\end{equation}
Thus the computation of the moment
$\mathbb E S_t^{(1-\beta)p}$ is reduced to the solution of the
one-dimensional killed parabolic problem \eqref{eq:pde-moment-vp}.
\end{cor}

We now turn to a distributional characterization of the SABR model based on
the functional representation established in Theorem~\ref{T1}. The key
ingredient is the affine dependence of the drift coefficient
$\mu_p$ in \eqref{definicja sigma i mu} on the Mellin parameter $p$.
Since the transformation $b_p=h_p\circ\varphi^{-1}$ is linear in $\mu_p$,
the function $b_p$ is affine in $p$ as well. More precisely, there exist
deterministic functions $\gamma_1$ and $\gamma_2$, independent of $p$, such
that
\[
b_p(x)=p\,\gamma_1(x)+\gamma_2(x),
\qquad
x\in(-\infty,\gamma).
\]
The functions $\gamma_1$ and $\gamma_2$ are given explicitly by
\begin{align}
\label{wzor na gamma1}
\gamma_1(x)
&:=
\frac{\alpha^2\varphi^{-1}(x)
-\alpha\rho(\varphi^{-1}(x))^2}
{\sigma(\varphi^{-1}(x))},
\\
\label{wzor na gamma2}
\gamma_2(x)
&:=
b_p(x)-p\,\gamma_1(x),
\end{align}
for $x\in(-\infty,\gamma)$.

This decomposition allows the Mellin parameter to be separated from the
remaining path functional in the representation of
Theorem~\ref{T1}. Consequently, the Mellin transform can be inverted
explicitly, leading to a distributional representation of $S_t$ in terms of
Brownian functionals. The resulting formula is stated in
Theorem~\ref{T22}.
 
\begin{tw} \label{T22} Let $\rho \in (-1,1)$. Then, for every $t>0$, we have
\begin{align}\label{wzor 1 tw o rozkladzie}
S_t^{1-\beta}\overset{d}{=} &\frac{(1-\beta)e^{-\frac{\alpha^2 t}{2}}}{\varphi^{-1}(u_0+W_t)}e^{V_{\int_{0}^{t}(\alpha^2 - \gamma^2_1(u_0+W_s))ds}}\notag\\[1ex]
&\times  e^{\int_{0}^{t}\gamma_1(u_0+W_s)dW_s-\int_{0}^{t}\gamma_1(u_0+W_s)\gamma_2(u_0+W_s)ds}\mathbf{1}_{\{t<\sigma^W_{\gamma-u_0}\}},
\end{align}
where the law of the process $W$ is characterized by 
\begin{align}\label{wzor 2 tw o rozkladzie}
	\mathbb{P}(W_{\cdot} \in \Delta, t < \sigma^W_{\gamma-u_0}) = \mathbb{E}e^{\int_0^t\gamma_2(u_0+B_s)dB_s - \frac12\int_0^t\gamma_2^2(u_0+B_s)ds}1_{\{B_{\cdot}\in \Delta, t<\sigma^{B}_{\gamma-u_0}\}},
\end{align}
for every Borel set
$\Delta\subset\mathcal C([0,t])$, where $B$ and $V$ are independent standard Brownian motions under
$\mathbb P$, and $\sigma_a^B$ denotes the first hitting time of the
level $a$ by $B$.

\end{tw}

In the regime $\rho\in[-1,1)$, the distribution of $S_t$ admits the
following alternative functional representation.
To this end, define the auxiliary functions
\[
\nu_1,\nu_2:(0,\infty)\to\mathbb{R}
\]
by
\begin{align*}
\nu_1(x)
&=
\frac{\alpha^2x-\alpha\rho x^2}{\sigma(x)},
\\
\nu_2(x)
&=
\frac{1}{\sigma(x)}
\left(
\frac{2-\beta}{2(1-\beta)}x^3
-\alpha\rho x^2
\right),
\qquad x>0.
\end{align*}

\begin{tw}\label{T222}
Let $\rho\in[-1,1)$. Then, for every $t>0$,
\[
S_t^{1-\beta}
\overset{d}{=}
(1-\beta)
e^{-\frac{\alpha^2t}{2}}
\exp\!\left(
\int_0^t
\nu_1(\theta_s^o)\,dW_s^o
+
V_{\int_0^t
(\alpha^2-\nu_1^2(\theta_s^o))\,ds}
\right)
(\theta_t^o)^{-1}
\mathbf 1_{\{t<\sigma^o\}},
\]
where $W^o$ and $V$ are independent standard Brownian motions, and
$\theta^o$ is the (weak) solution of
\[
d\theta_t^o
=
\sigma(\theta_t^o)\,dW_t^o
+
\sigma(\theta_t^o)\nu_2(\theta_t^o)\,dt,
\qquad
t<\sigma^o,
\]
with $\sigma^o$ denoting the explosion time of $\theta^o$.
\end{tw}

\subsection{Financial consequences and numerical implementation}
Theorem~\ref{T222} expresses the SABR asset price as an explicit
functional of a one-dimensional diffusion and two independent Brownian
motions. In particular, it eliminates the moment-dependent change of
measure appearing in the previous representation and forms the basis
for the option pricing results developed below.

\begin{cor}[Black--Scholes mixture representation]
\label{cor:mixture-bs}
Let $\beta\in[0,1)$ and set
\[
q:=1-\beta.
\]
As a consequence of Theorem~\ref{T222}, the undiscounted European call
price
\[
C(t,K):=\mathbb E\big[(S_t-K)^+\big]
\]
admits a Black--Scholes mixture representation.
Define
\[
A_t
:=
\log(qy_0)
-\frac{\alpha^2t}{2}
+\int_0^t\nu_1(\theta_s^o)\,dW_s^o
-\log\theta_t^o,
\]
and
\[
\Sigma_t^2
:=
\int_0^t
\bigl(\alpha^2-\nu_1^2(\theta_s^o)\bigr)\,ds.
\]
Furthermore, set
\[
\mathcal V_t
:=
\frac{\Sigma_t}{q},
\qquad
\mathcal S_t
:=
\exp\!\left(
\frac{A_t}{q}
+
\frac{\Sigma_t^2}{2q^2}
\right).
\]
Then
\begin{gather}\label{Option price representation SABR}
C(t,K)
=
\mathbb E\!\left[
\mathbf 1_{\{t<\sigma^o\}}
\,BS(\mathcal S_t,K,\mathcal V_t)
\right],
\end{gather}
where
\[
BS(S,K,\nu)
=
S\Phi(d_1)-K\Phi(d_2),
\]
with
\[
d_1
=
\frac{\log(S/K)}{\nu}
+\frac{\nu}{2},
\qquad
d_2
=
d_1-\nu,
\]
and $\Phi$ denotes the standard normal distribution function.

If $\nu=0$, the Black--Scholes expression is understood in the limiting
sense,
\[
BS(S,K,0)=(S-K)^+.
\]
\end{cor}

The representation \eqref{Option price representation SABR} has a direct numerical interpretation. Rather than simulating the full two-dimensional SABR system and evaluating the payoff pathwise, the remaining independent Gaussian component in the representation of Theorem \ref{T222} is integrated out analytically through the Black - Scholes formula. This leads naturally to a Rao-Blackwellized Monte Carlo estimator. More precisely, let $\mathcal G_T$ denote the $\sigma$-field generated by the path of $\theta^o$ up to time $T$. The Black-Scholes random variable appearing in Corollary \ref{cor:mixture-bs} is the conditional expectation of the corresponding SABR payoff given $\mathcal G_T$. Hence, by the law of total variance,
\[
    \operatorname{Var}
    \!\left(
        \mathbb E
        \!\left[
            (S_T-K)^+
            \mid
            \mathcal G_T
        \right]
    \right)
    \le
    \operatorname{Var}
    \!\left(
        (S_T-K)^+
    \right),
\]
Thus the resulting estimator is unbiased and has variance no larger than the standard pathwise Monte Carlo estimator.
The resulting Monte Carlo procedure for computing European call prices can be summarized as follows.

\begin{enumerate}
    \item Simulate the one-dimensional diffusion $\theta^o$ up to
    maturity $t$.

    \item Along each simulated path compute
    \[
        A_t
        =
        \log(qy_0)
        -\frac{\alpha^2t}{2}
        +
        \int_0^t
        \nu_1(\theta_s^o)\,dW_s^o
        -
        \log\theta_t^o
    \]
    and
    \[
        \Sigma_t^2
        =
        \int_0^t
        \bigl(\alpha^2-\nu_1^2(\theta_s^o)\bigr)\,ds.
    \]

    \item If the path reaches the explosion time before maturity,
    i.e. $t\geq\sigma^o$, assign payoff zero.

    \item Otherwise evaluate the Black--Scholes price
    \[
        BS(\mathcal S_t,K,\mathcal V_t),
    \]
    where
    \[
        \mathcal S_t
        =
        \exp\!\left(
            \frac{A_t}{q}
            +
            \frac{\Sigma_t^2}{2q^2}
        \right),
        \qquad
        \mathcal V_t
        =
        \frac{\Sigma_t}{q}.
    \]

    \item Average the resulting values over all simulated paths.
\end{enumerate}

The same option price representation also leads to a characterization of the Black-Scholes implied volatility, as described in the following remark.

\begin{rem}[Implied-volatility representation]
\label{rem:impvol}
Assume that
\[
(S_0-K)^+
<
C_{\mathrm{SABR}}(T,K)
<
S_0.
\]
Since the Black--Scholes price is a continuous strictly increasing
function of the volatility parameter, the implied volatility is uniquely
determined by
\[
BS(S_0,K,\sigma_{\mathrm{impl}}(T,K)\sqrt T)
=
\mathbb E\!\left[
\mathbf 1_{\{T<\sigma^o\}}
BS(\mathcal S_T,K,\mathcal V_T)
\right].
\]
Equivalently,
\[
\sigma_{\mathrm{impl}}(T,K)
=
\frac1{\sqrt T}
BS^{-1}_{S_0,K}
\!\left(
\mathbb E\!\left[
\mathbf 1_{\{T<\sigma^o\}}
BS(\mathcal S_T,K,\mathcal V_T)
\right]
\right),
\]
where $BS^{-1}_{S_0,K}$ denotes inversion of the Black--Scholes pricing
formula with respect to total volatility.
\end{rem}

Although this representation does not yield a closed-form expression for the implied volatility, it reduces its computation to the inversion of a Black–Scholes price obtained by averaging over the auxiliary diffusion. It therefore provides an exact probabilistic benchmark complementary to the classical asymptotic approximation of Hagan et al. Whereas Hagan's formula gives a fast explicit approximation particularly suitable for calibration, the present representation is exact and applies to arbitrary maturities and strikes. It can thus be used, in particular, to assess the accuracy of asymptotic implied-volatility formulas.

A systematic numerical comparison with standard Monte Carlo methods and with Hagan's implied-volatility approximation is beyond the scope of the present work, but the representation developed above provides a natural framework for such comparisons.

\section{Proofs}\label{Section Proofs} 

\subsection{ Proof of Theorem ~\ref{thm:main}}
The fact that any moment of positive order of $S$ is finite follows from Remark 5.2 in \cite{AndersenPiterbag}. 
Fix $p>0$ and define the measure $\QQ=\QQ^{(p)}$ via the density:
\begin{align}\label{Q density}
    \frac{\dd \QQ}{\dd \PP}|_{\mathcal{F}_t} 
    = 
    e^{p\alpha Z_t - \frac{p^2\alpha^2}{2}t},
    \qquad t\ge 0.
\end{align}
It follows from \eqref{eq:theta-def}, for $t\ge0$, that
\begin{align*}
    \EE S_t^{(1-\beta)p}
    &= (1-\beta)^p
    \EE Y_t^p \theta^{-p}_t
    \\&=(1-\beta)^p
    \EE e^{p\alpha Z_t - \frac{p\alpha^2 t}{2}} \theta^{-p}_t\\ 
    &= (1-\beta)^p\cdot e^{\frac{1}{2}\alpha^2 p(p-1)t}
    \EE e^{p\alpha Z_t - \frac{p^2\alpha^2 t}{2}} \theta^{-p}_t\\
    &=(1-\beta)^p \cdot e^{\frac{1}{2}\alpha^2 p(p-1)t}
    \EE_\QQ \theta^{-p}_t.
\end{align*}
We now derive the dynamics of $\theta$ under $\mathbb{Q}$.
Since $\theta_t=f(S_t,Y_t)$ with $f(x,y):=(1-\beta)x^{\beta-1}y$ and $f$ is a $C^2$-function on the set $(0,+\infty)\times(0,+\infty)$,
we can apply the It\^o formula before $S$ hits zero. This yields
\begin{align}\label{proof theorem main}\nonumber
    d\theta_t =&-(1-\beta)^2Y_tS_t^{\beta-2} dS_t+(1-\beta)S_t^{\beta-1} dY_t\\ \nonumber
    &-\frac{1}{2}(1-\beta)^2(\beta-2)Y_tS_t^{\beta-3}d\langle S\rangle_t-2\cdot\frac{1}{2}(1-\beta)^2S_t^{\beta-2}d\langle S,Y\rangle_t\\\nonumber
    =&-(1-\beta)^2Y_t^2S_t^{2\beta-2}dW_t+\alpha(1-\beta)Y_tS_t^{\beta-1}dZ_t\\ \nonumber
    &-\frac{1}{2}(1-\beta)^2(\beta-2)Y_t^3S_t^{3\beta-3}dt-\alpha\rho(1-\beta)^2Y_t^2S_t^{2\beta-2}dt\\
    =&-\theta_t^2dW_t+\alpha\theta_tdZ_t+\frac{2-\beta}{2-2\beta}\theta_t^3dt-\alpha\rho\theta_t^2dt, \qquad t>0.
    \end{align}
Since $V$ and $Z$ are independent, the change of measure~\eqref{Q density} does not affect $V$.
Thus, by the Girsanov theorem, the process $B_s = Z_s -\alpha ps$ is a Brownian motion under $\QQ$, independent of $V$. Using $V$ and $B$ 
in \eqref{proof theorem main} yields
\begin{align*}
    d\theta_t =&-\theta_t^2\Big(\alpha p\rho dt+\rho dB_t+\sqrt{1-\rho^2}dV_t\Big)+\alpha\theta_t\Big(dB_t+\alpha pdt\Big)\\
		& \qquad \qquad \qquad \qquad \qquad \qquad \qquad +\frac{2-\beta}{2-2\beta}\theta_t^3dt-\alpha\rho\theta_t^2dt\\
    =&\mu_p(\theta_t)dt+\theta_t(\alpha-\rho\theta_t)dB_t-\theta_t^2\sqrt{1-\rho^2}dV_t,
\end{align*}
where 
$$
\mu_p(x):=\alpha^2 p x-\rho\alpha (p+1)x^2+\frac{2-\beta}{2-2\beta}x^3, \quad x\geq 0.
$$
It follows that the generator of $\theta$ takes the form, see, e.g., \cite[p. 281-283]{KS}. 
\begin{equation}\label{generator - postac}
    \mathcal{A}f(x)
    = \frac{1}{2}
    \Big(x^4-2\alpha\rho x^3+\alpha^2x^2\Big)
    f^{\prime\prime}(x)+\mu_p(x) f^\prime(x), \qquad x\geq0,
\end{equation}
for suitably regular functions $f\in\mathcal{C}^2((0,+\infty))$.
Note that $\mathcal{A}$ defined in~\eqref{generator - postac} is also a generator of the equation 
\begin{gather}\label{rownanie na theta 1 dim}
    d\theta_t 
    = 
    \sqrt{\theta_t^4-2\alpha\rho \theta_t^3+\alpha^2\theta_t^2}dB_t
    +
    \mu_p(\theta_t) dt;
    \quad t\ge 0,
\end{gather}
driven by a single $\mathbb{Q}$-Wiener process $B$. Since the one-dimensional diffusion
\eqref{rownanie na theta 1 dim}
has generator \eqref{generator - postac},
the claimed characterization follows from the standard
equivalence between weak solutions and martingale problems
(see, e.g., \cite{KS}).
\hfill$\square$

\subsection{ Proof of Proposition \ref{prop:explicit-bp}}
Since
\[
\varphi'(x)\sigma(x)=1,
\]
differentiation yields
\[
\varphi''(x)
=
-\frac{\sigma'(x)}{\sigma^{2}(x)},
\]
and therefore
\[
h_p(x)
=
\frac12
\varphi''(x)\sigma^{2}(x)
+
\frac{\mu_p(x)}{\sigma(x)}
=
-\frac12\sigma'(x)
+
\frac{\mu_p(x)}{\sigma(x)}.
\]
Now
\[
\sigma(x)
=
x\sqrt{x^{2}-2\alpha\rho x+\alpha^{2}},
\]
so that
\[
\sigma'(x)
=
\frac{
2x^{2}
-
3\alpha\rho x
+
\alpha^{2}
}
{
\sqrt{x^{2}-2\alpha\rho x+\alpha^{2}}
}.
\]
Consequently,
\[
h_p(x)
=
\frac{
\displaystyle
\frac{\beta}{2(1-\beta)}x^{2}
+
\alpha\rho
\left(
\frac12-p
\right)x
+
\alpha^{2}
\left(
p-\frac12
\right)
}
{
\sqrt{x^{2}-2\alpha\rho x+\alpha^{2}}
}.
\]
Next, solving explicitly the equation
\[
\varphi(x)=u
\]
gives
\[
\varphi^{-1}(u)
=
\frac{\alpha}
{\rho-\delta\sinh u},
\qquad
\delta=\sqrt{1-\rho^{2}}.
\]
Substituting this expression into the above formula for
$h_p$
yields
\[
b_p(u)
=
\alpha
\left[
\frac{\beta}
{2(1-\beta)\,
\delta\cosh u\,
(\rho-\delta\sinh u)}
-
\left(
p-\frac12
\right)
\tanh u
\right].
\]
The case $\beta=0$ follows immediately, and the bound
\[
|b_p(u)|
\le
\alpha
\left|
p-\frac12
\right|
\]
is a direct consequence of
$|\tanh u|\le1$.
\hfill$\square$

\subsection{Proof of Theorem \ref{T1}}

We start with an auxiliary result.
\begin{prop}\label{prop charakteryzacja theta} 

Let $\rho\in[-1,1)$. Let $\varphi$ be a solution of the equation 
\begin{gather}\label{fi}
\varphi^{\prime}(x)\cdot \sigma(x)=1,  \ \varphi(x_0)=\varphi_0,\quad x_0,\varphi_0>0, \quad x>0.
\end{gather}
Define
\begin{gather}\label{h}
h_p(x):=\frac{1}{2}\varphi^{\prime\prime}(x)\sigma^2(x)+\frac{\mu_p(x)}{\sigma(x)}, \quad x>0.
\end{gather}
Then the process $\theta$ solves \eqref{rownanie na teta} if and only if it solves the following pathwise differential equation
\begin{gather}\label{pierwszy wzor}
\theta_t=\varphi^{-1}\left(\varphi(\theta_0)+B_t+\int_{0}^{t}h_p(\theta_s)\dd s\right), \quad t\geq 0.
\end{gather}

\end{prop}

\noindent
{\bf Proof:}  It follows from \eqref{fi} that
\begin{gather}\label{wzory na pochodne fi}
\varphi^{\prime}(x)=\frac{1}{\sigma(x)}, \quad \varphi^{\prime\prime}(x)=-\frac{\sigma^\prime(x)}{\sigma^2(x)}=
-\frac{x\left(2x^2 - 3\alpha\rho x + \alpha^2\right)}{\left(x^4 - 2\alpha\rho x^3 + \alpha^2x^2\right)^{3/2}}, \quad x>0,
\end{gather}
so both functions $\varphi^\prime,\varphi^{\prime\prime}$ are continuous in $(0,+\infty)$. Since $\varphi^{\prime}$ is strictly positive,  $\varphi^{-1}$ exists and is differentiable and for $x>0$
\begin{align}\label{wzory na pochodne odwrotnej do fi}
(\varphi^{-1})^{\prime}(x) &=\frac{1}{\varphi^\prime(\varphi^{-1}(x))}=\sigma(\varphi^{-1}(x)),\notag\\
 (\varphi^{-1})^{\prime\prime}(x) &=\sigma^\prime(\varphi^{-1}(x))\cdot \sigma(\varphi^{-1}(x)).
\end{align}
Hence $\varphi^{-1}$ is also a $C^2$-class function.

Let $\theta$ be a solution of \eqref{rownanie na teta}. By the It\^o formula $\varphi(\theta)$ admits the following representation on the set before $\theta$ explodes:
\begin{align}\label{basic formula}\nonumber 
\varphi(\theta_t)&=\varphi(\theta_0)+\int_{0}^{t}\varphi^\prime(\theta_s)\Big(\sigma(\theta_s)\dd B_s+\mu_p(\theta_s) \dd s\Big)+\frac{1}{2}\varphi^{\prime\prime}(\theta_s)\sigma^2(\theta_s) \dd s\\ \nonumber
&=\varphi(\theta_0)+B_t+\int_{0}^{t}\left(\frac{\mu_p(\theta_s)}{\sigma(\theta_s)}+\frac{1}{2}\varphi^{\prime\prime}(\theta_s)\sigma^2(\theta_s)\right) \dd s\\
&=\varphi(\theta_0)+B_t+\int_{0}^{t}h_p(\theta_s)\dd s, \qquad t\geq 0,
\end{align}
so \eqref{pierwszy wzor} follows. If $\theta$ solves \eqref{pierwszy wzor} then 
$$
\theta_t=\varphi^{-1}(H_t), \quad \text{with} \quad H_t:=\varphi(\theta_0)+B_t+\int_{0}^{t}h_p(\theta_s)ds, \quad t\geq 0.
$$ Then, by It\^o's formula, \eqref{wzory na pochodne odwrotnej do fi} and \eqref{wzory na pochodne fi}, one easily obtains \eqref{rownanie na teta}. \hfill$\square$

\vskip2ex
\noindent
{\bf Proof of Theorem \ref{T1}:} $(1.)$ By Theorem~3.2 in \cite{KaratzasRuf}, for $p>0$,
\[
\mathbb{E}^{(p)}\left[\theta_t^{-p}\mathbf 1_{\{t<\sigma\}}\right]
=
\mathbb{E}^{o}\left[
e^{\int_0^t\nu_p(\theta_s^o)dB_s^o
-\frac12\int_0^t\nu_p^2(\theta_s^o)ds}
(\theta_t^o)^{-p}\mathbf 1_{\{t<\sigma^o\}}
\right],
\]
where $\nu_p(x):=\mu_p(x)/\sigma(x)$.
Combining this with \eqref{rownanie na momenty} gives
\eqref{firstexplo}.

We now prove \eqref{Lampertiii}. Observe that \eqref{theta trifftless} may be
written in the form
\[
d\theta_t^o
=
\theta_t^o\,\sigma_0(\theta_t^o)\,dB_t^o,
\qquad
\sigma_0(x):=\sqrt{x^2-2\alpha\rho x+\alpha^2},
\quad x>0.
\]
Applying It\^o's formula to $(\theta_t^o)^2$ yields
\[
d(\theta_t^o)^2
=
2(\theta_t^o)^2\sigma_0(\theta_t^o)\,dB_t^o
+
(\theta_t^o)^2\sigma_0^2(\theta_t^o)\,dt,
\]
or equivalently,
\begin{equation}\label{wwwzzor na ilorazz}
\frac{d(\theta_t^o)^2}{2(\theta_t^o)^2}
=
\sigma_0(\theta_t^o)\,dB_t^o
+
\frac12\sigma_0^2(\theta_t^o)\,dt,
\qquad
t<\sigma^o.
\end{equation}
Introduce the increasing time change
\[
\tau_t
:=
\inf\left\{
s<\sigma^o:
\int_0^s
\sigma_0^2(\theta_u^o)\,du
\ge t
\right\},
\qquad t\ge0.
\]
By the Dambis--Dubins--Schwarz theorem,
\[
W_t
:=
\int_0^{\tau_t}
\sigma_0(\theta_u^o)\,dB_u^o
\]
is a standard Brownian motion under $\mathbb Q^o$.
Therefore \eqref{wwwzzor na ilorazz} becomes
\[
\frac{d(\theta_{\tau_t}^o)^2}
     {2(\theta_{\tau_t}^o)^2}
=
dW_t+\frac12\,dt.
\]
Applying It\^o's formula to
$\log (\theta_{\tau_t}^o)^2$
gives
\[
d\log (\theta_{\tau_t}^o)^2
=
2\,dW_t-dt,
\]
and hence
\[
\log (\theta_{\tau_t}^o)^2
=
\log \theta_0^2
+
2W_t
-
t.
\]
Consequently,
\begin{equation}\label{theta zero kwadrat representacja}
(\theta_{\tau_t}^o)^2
=
\theta_0^2
e^{2W_t-t},
\qquad
t\ge0,
\end{equation}
or equivalently,
\[
\theta_{\tau_t}^o
=
\theta_0
e^{W_t-\frac t2}.
\]
Finally, since
\[
t
=
\int_0^{\tau_t}
\sigma_0^2(\theta_u^o)\,du,
\]
differentiation with respect to $t$ yields
\[
d\tau_t
=
\frac{dt}
{\sigma_0^2(\theta_{\tau_t}^o)}.
\]
Using the representation above, we obtain
\[
d\tau_t
=
\frac{dt}
{\sigma_0^2\!\left(
\theta_0e^{W_t-\frac t2}
\right)},
\]
and therefore
\begin{equation}\label{tau t reprezentacja}
\tau_t
=
\int_0^t
\frac{1}
{\sigma_0^2\!\left(
\theta_0e^{W_s-\frac s2}
\right)}
\,ds.
\end{equation}
Combining \eqref{theta zero kwadrat representacja} and
\eqref{tau t reprezentacja} yields the Lamperti-type representation
\[
\theta^o_{
\int_0^t
\frac{1}
{\sigma_0^2\!\left(
\theta_0e^{W_s-\frac s2}
\right)}
\,ds
}
=
\theta_0
e^{W_t-\frac t2},
\qquad
t<\sigma^o,
\]
which is precisely \eqref{Lampertiii}.

\noindent
$(2.)$ Since $\varphi$ given by \eqref{varphi wzooorr} is strictly increasing,
for every $p>0$ we have the layer-cake representation
\begin{align}\label{zaba}
\mathbb E^{(p)}[\theta_t^{-p}]
=
p\int_0^\infty
v^{p-1}
\mathbb Q^{(p)}
\!\left(
\theta_t<\frac1v
\right)
\,dv
=
p\int_0^\infty
v^{p-1}
\mathbb Q^{(p)}
\!\left(
\varphi(\theta_t)
<
\varphi\!\left(\frac1v\right)
\right)
\,dv.
\end{align}
Recall from Proposition~\ref{prop charakteryzacja theta} that
\[
U_t:=\varphi(\theta_t)
\]
satisfies
\begin{equation}\label{proces U}
U_t
=
u_0+B_t+\int_0^tb_p(U_s)\,ds,
\qquad
u_0
=
\varphi(\theta_0),
\end{equation}
up to the explosion time of $\theta$.

Moreover,
\[
\lim_{x\to\infty}\varphi(x)
=
\gamma
:=
\frac12
\log\frac{1+\rho}{1-\rho},
\qquad
\rho\in(-1,1),
\]
so that explosion of $\theta$ corresponds precisely to the first hitting of
the level $\gamma$ by $U$. Hence
\[
\sigma
=
\sigma_\gamma^U
:=
\inf\{t>0:U_t=\gamma\}.
\]
The driftless counterpart of \eqref{proces U} is simply
\[
U_t^o=u_0+B_t^o.
\]
Applying Theorem~3.2 of \cite{KaratzasRuf} therefore yields
\begin{align*}
&\mathbb Q^{(p)}
\!\left(
U_t<\varphi(1/v),
\,t<\sigma_\gamma^U
\right)
\\
&\qquad
=
\mathbb E^o
\Bigg[
e^{\int_0^tb_p(U_s^o)\,dB_s^o
-\frac12\int_0^tb_p^2(U_s^o)\,ds}
\,
\mathbf1_{\{
U_t^o<\varphi(1/v)
\}}
\,
\mathbf1_{\{
t<\sigma_\gamma^{U^o}
\}}
\Bigg].
\end{align*}

Since
\[
U_t^o=u_0+B_t^o,
\qquad
\sigma_\gamma^{U^o}
=
\sigma_{\gamma-u_0}^{B^o},
\]
and
\[
U_t^o<\varphi(1/v)
\iff
v<
\frac1{\varphi^{-1}(u_0+B_t^o)},
\]
formula \eqref{zaba} becomes
\begin{align*}
\mathbb E^{(p)}[\theta_t^{-p}]
&=
p
\int_0^\infty
v^{p-1}
\\
&\qquad\times
\mathbb E^o
\Bigg[
e^{\int_0^tb_p(u_0+B_s^o)\,dB_s^o
-\frac12\int_0^tb_p^2(u_0+B_s^o)\,ds}
\,
\mathbf1_{\{
t<\sigma_{\gamma-u_0}^{B^o}
\}}
\,
\mathbf1_{\{
v<
(\varphi^{-1}(u_0+B_t^o))^{-1}
\}}
\Bigg]
\,dv.
\end{align*}
Fubini's theorem now yields
\[
\mathbb E^{(p)}[\theta_t^{-p}]
=
\mathbb E^o
\left[
e^{\int_0^tb_p(u_0+B_s^o)\,dB_s^o
-\frac12\int_0^tb_p^2(u_0+B_s^o)\,ds}
\,
\mathbf1_{\{
t<\sigma_{\gamma-u_0}^{B^o}
\}}
\,
\frac1{(\varphi^{-1}(u_0+B_t^o))^p}
\right],
\]
which proves \eqref{wzor w T1}.
Finally,
\[
\sigma=\sigma_\gamma^U,
\]
and another application of the generalized Girsanov theorem gives
\begin{align*}
\mathbb Q^{(p)}(\sigma>t)
&=
\mathbb Q^{(p)}(\sigma_\gamma^U>t)
\\
&=
\mathbb E^o
\left[
e^{\int_0^tb_p(U_s^o)\,dB_s^o
-\frac12\int_0^tb_p^2(U_s^o)\,ds}
\,
\mathbf1_{\{
t<\sigma_\gamma^{U^o}
\}}
\right]
\\
&=
\mathbb E^o
\left[
e^{\int_0^tb_p(u_0+B_s^o)\,dB_s^o
-\frac12\int_0^tb_p^2(u_0+B_s^o)\,ds}
\,
\mathbf1_{\{
t<\sigma_{\gamma-u_0}^{B^o}
\}}
\right],
\end{align*}
which is precisely \eqref{rozklad dijscia do zera przez S}.

\ \\
$(3.)$ We first prove that there exists $p>0$ such that
\begin{equation}\label{finite explosion time}
\mathbb{Q}^{(p)}(\sigma<+\infty)=1.
\end{equation}

Recall that $\sigma$ coincides with the explosion time of the auxiliary
diffusion $\theta$. Hence it suffices to show that $\theta$ explodes almost
surely under $\mathbb{Q}^{(p)}$. To this end we apply Feller's test for
explosions (see, e.g., \cite{KS,KaratzasRuf}) on the state space
$I=(0,+\infty)$.

The diffusion coefficient and drift of $\theta$ are given by
\begin{gather}\label{s and b}
\sigma(x)
=
\sqrt{x^4-2\alpha\rho x^3+\alpha^2x^2},
\qquad
\mu_p(x)
=
\alpha^2px
-\alpha\rho(p+1)x^2
+\frac{2-\beta}{2-2\beta}x^3,
\end{gather}
for $x>0$.

Since $\rho\in[-1,1)$, the function $\sigma$ has a unique zero at the
origin and
\[
\frac{1+|\mu_p(x)|}{\sigma^2(x)}
\]
is locally integrable on $(0,+\infty)$, so the assumptions of Feller's test
are satisfied.

Introduce the scale function
\[
p(x)
:=
\int_{c_0}^{x}
e^{-2F(y)}\,dy,
\]
where
\[
F(x)
:=
\int_{c_0}^{x}
f(y)\,dy,
\qquad
f(x)
:=
\frac{\mu_p(x)}{\sigma^2(x)},
\]
and the corresponding Feller test function
\begin{equation}\label{Feller test function}
v(x)
=
\int_{c_0}^{x}
\frac{e^{2F(z)}}{\sigma^2(z)}
\left(
\int_{z}^{x}
e^{-2F(y)}\,dy
\right)
dz.
\end{equation}

We shall show that, for a suitable choice of $p>0$,
\[
v(+\infty)<+\infty,
\qquad
p(0+)=-\infty.
\]
By Proposition~5.32(ii) in \cite{KS}, this implies
\eqref{finite explosion time}.

We first verify that $v(+\infty)<+\infty$. Since
\begin{equation}\label{f pomocniczy}
f(x)
=
\frac{1}{x}
\cdot
\frac{\alpha^2px-\alpha\rho(p+1)x^2+
\frac{2-\beta}{2-2\beta}x^3}
{x^3-2\alpha\rho x^2+\alpha^2x}
=
\frac{h(x)}{x},
\end{equation}
and
\[
\lim_{x\to+\infty}h(x)
=
\frac{2-\beta}{2-2\beta}
=:C,
\]
for every $\varepsilon>0$ one can choose $c_0=c_0(\varepsilon)$ sufficiently
large so that
\[
C-\varepsilon
\le
h(x)
\le
C+\varepsilon,
\qquad
x\ge c_0.
\]
Hence
\[
(C-\varepsilon)
(\log x-\log c_0)
\le
F(x)
\le
(C+\varepsilon)
(\log x-\log c_0),
\qquad
x\ge c_0.
\]

Consequently,
\begin{equation}\label{oszacowania w}
v(x)
\le
\bar v_\varepsilon(x),
\qquad
x\ge c_0,
\end{equation}
where
\[
\bar v_\varepsilon(x)
=
\int_{c_0}^{x}
\frac{
e^{2(C+\varepsilon)(\log z-\log c_0)}
}
{z^4-2\alpha\rho z^3+\alpha^2z^2}
\left(
\int_z^x
e^{-2(C-\varepsilon)(\log y-\log c_0)}
\,dy
\right)
dz.
\]

Since $C\ge1$, choose $\varepsilon>0$ such that
$C-\varepsilon>\frac12$. Then
\begin{align*}
\bar v_\varepsilon(x)
&=
\frac{e^{-4\varepsilon\log c_0}}
{1-2(C-\varepsilon)}
\Bigg[
x^{1-2(C-\varepsilon)}
\int_{c_0}^{x}
\frac{dz}
{z^{-2(C+\varepsilon)}
(z^4-2\alpha\rho z^3+\alpha^2z^2)}
\\
&\qquad\qquad
-
\int_{c_0}^{x}
\frac{dz}
{z^{-4\varepsilon}
(z^3-2\alpha\rho z^2+\alpha^2z)}
\Bigg].
\end{align*}

The second integral converges at $+\infty$. If
$C+\varepsilon<\frac32$, then the first integral also converges and
$x^{1-2(C-\varepsilon)}\to0$, implying
$\bar v_\varepsilon(+\infty)<+\infty$.

If instead
$C+\varepsilon\ge\frac32$, L'Hospital's rule gives
\begin{align*}
&\lim_{x\to+\infty}
x^{1-2(C-\varepsilon)}
\int_{c_0}^{x}
\frac{dz}
{z^{-2(C+\varepsilon)}
(z^4-2\alpha\rho z^3+\alpha^2z^2)}
\\
&=
\frac{1}{2(C-\varepsilon)-1}
\lim_{x\to+\infty}
\frac{1}
{x^{-4\varepsilon}
(x^2-2\alpha\rho x+\alpha^2)}
=
0.
\end{align*}
Hence also in this case
$\bar v_\varepsilon(+\infty)<+\infty$.
By \eqref{oszacowania w},
\[
v(+\infty)<+\infty.
\]

It remains to verify that
$p(0+)=-\infty$.
Since
\[
\lim_{x\downarrow0}h(x)=p,
\]
for every $\varepsilon>0$ there exists
$c_0>0$ such that
\[
\frac{p-\varepsilon}{x}
\le
f(x)
\le
\frac{p+\varepsilon}{x},
\qquad
x\in(0,c_0).
\]
Therefore
\[
-(p+\varepsilon)
(\log c_0-\log x)
\le
F(x)
\le
-(p-\varepsilon)
(\log c_0-\log x),
\]
and
\begin{align*}
-p(x)
=
\int_x^{c_0}
e^{-2F(y)}\,dy
&\ge
\int_x^{c_0}
e^{2(p-\varepsilon)
\log(c_0/y)}
\,dy
\\
&=
\int_x^{c_0}
\left(
\frac{c_0}{y}
\right)^{2(p-\varepsilon)}
dy.
\end{align*}
Choosing $p$ so that
$2(p-\varepsilon)>1$, the last integral diverges as
$x\downarrow0$, and therefore
\[
p(0+)=-\infty.
\]

Consequently,
\[
v(+\infty)<+\infty,
\qquad
p(0+)=-\infty,
\]
and Feller's test yields
\[
\mathbb Q^{(p)}(\sigma<+\infty)=1.
\]

Finally, we transfer this conclusion to the original measure
$\mathbb P$. Since $\mathbb Q^{(p)}$ and $\mathbb P$ are equivalent on
every $\mathcal F_t$, we have for every $n\in\mathbb N$,
\[
\mathbb Q^{(p)}(\sigma\le n)=0
\quad\Longleftrightarrow\quad
\mathbb P(\sigma\le n)=0.
\]
If $\mathbb P(\sigma<+\infty)=0$, then
$\mathbb P(\sigma\le n)=0$ for every $n$, and hence also
$\mathbb Q^{(p)}(\sigma\le n)=0$ for every $n$. Consequently,
\[
\mathbb Q^{(p)}(\sigma<+\infty)
=
\lim_{n\to\infty}
\mathbb Q^{(p)}(\sigma\le n)
=
0,
\]
contradicting the already established identity
$\mathbb Q^{(p)}(\sigma<+\infty)=1$.
Therefore
\[
\mathbb P(\sigma<+\infty)>0,
\]
which proves \eqref{S reaches zero in finite time}.
\hfill$\square$

\subsection{ Proof of Corollary \ref{cor:one-dimensional-PDE}}
By the Brownian functional representation \eqref{wzor w T1},
\[
\mathbb E
\left(
    \frac{S_t^{(1-\beta)p}}{(1-\beta)^p}
\right)
=
e^{\frac{\alpha^2p(p-1)t}{2}}
\mathbb E
\left[
    \mathcal E_t
    \frac{\mathbf 1_{\{t<\sigma^W_{\gamma-u_0}\}}}
    {(\varphi^{-1}(u_0+W_t))^p}
\right],
\]
where
\[
    \mathcal E_t
    =
    \exp\left\{
        \int_0^t b_p(u_0+W_s)\,dW_s
        -
        \frac12
        \int_0^t b_p^2(u_0+W_s)\,ds
    \right\}.
\]
By Girsanov's theorem, the last expectation is equal to
\[
    \mathbb E_{u_0}
    \left[
        g_p(U_t)
        \mathbf 1_{\{t<\sigma_\gamma^U\}}
    \right]
    =
    v_p(t,u_0),
\]
where $U$ solves
\[
    dU_t=dB_t+b_p(U_t)\,dt,
    \qquad U_0=u_0,
\]
and is killed upon reaching $\gamma$.

Therefore
\[
\mathbb E
\left(
    \frac{S_t^{(1-\beta)p}}{(1-\beta)^p}
\right)
=
e^{\frac{\alpha^2p(p-1)t}{2}}
v_p(t,u_0),
\]
which gives \eqref{eq:moment-pde-representation}.

Finally, by the Feynman--Kac formula for killed diffusions, the function
\[
    v_p(t,u)
    =
    \mathbb E_u
    \left[
        g_p(U_t)
        \mathbf 1_{\{t<\sigma_\gamma^U\}}
    \right]
\]
solves the boundary value problem \eqref{eq:pde-moment-vp}. \hfill $\square$

\subsection{Proof of Theorem \ref{T22}}
We start with a lemma.

\begin{lem}\label{T2}
Let $\rho\in(-1,1)$ and let $V,\widetilde V$ be two standard Brownian
motions independent of $S$. Then, for every $t>0$,
\begin{align}\label{rownosc wg rozkladu complex}
\frac{S_t^{1-\beta}}{1-\beta}
e^{\frac{\alpha^2t}{2}+i\alpha V_t}
\overset{d}{=}
&
\frac{1}{\varphi^{-1}(u_0+W_t)}
\mathbf 1_{\{t<\sigma^W_{\gamma-u_0}\}}
\notag\\
&\times
\exp\Bigg(
i\alpha\widetilde V_t
+
V_{\int_0^t(\alpha^2-\gamma_1^2(u_0+W_s))\,ds}
\notag\\
&
\qquad\qquad
+
\int_0^t
\gamma_1(u_0+W_s)\,dW_s
-
\int_0^t
\gamma_1(u_0+W_s)\gamma_2(u_0+W_s)\,ds
\Bigg),
\end{align}
where $W$ is a (possibly weak) solution of
\[
\widetilde W_t
=
W_t
-
\int_0^t
\gamma_2(u_0+W_s)\,ds,
\qquad
t<\sigma^W_{\gamma-u_0},
\]
with $W_0=0$, and $\widetilde W$ is a Brownian motion independent of
$V$ and $\widetilde V$.
\end{lem}
\noindent
{\bf Proof:}
By independence of $S$ and $V$, together with
\eqref{wzor w T1}, we obtain for every $p>0$
\begin{align}
\label{wzor 1 w T2}
\mathbb E
\left[
\left(
\frac{S_t^{1-\beta}}{1-\beta}
e^{\frac{\alpha^2t}{2}+i\alpha V_t}
\right)^p
\right]
=
\mathbb E
\left[
e^{
\int_0^tb_p(u_0+W_s)\,dW_s
-
\frac12
\int_0^tb_p^2(u_0+W_s)\,ds
}
f_t(W)
\right],
\end{align}
where
\[
f_t(W)
=
\mathbf 1_{\{t<\sigma^W_{\gamma-u_0}\}}
\frac{1}
{\bigl(\varphi^{-1}(u_0+W_t)\bigr)^p}.
\]

Since $b_p$ depends affinely on $p$, namely
\begin{equation}\label{b dekompozycja}
b_p(x)
=
p\gamma_1(x)+\gamma_2(x),
\qquad
x\in(-\infty,\gamma),
\end{equation}
substituting \eqref{b dekompozycja} into
\eqref{wzor 1 w T2} yields
\begin{align}
\label{martp}
&
\mathbb E
\left[
\left(
\frac{S_t^{1-\beta}}{1-\beta}
e^{\frac{\alpha^2t}{2}+i\alpha V_t}
\right)^p
\right]
\notag\\
&=
\mathbb E
\Bigg[
e^{
\int_0^t
\gamma_2(u_0+W_s)\,dW_s
-
\frac12
\int_0^t
\gamma_2^2(u_0+W_s)\,ds
}
\notag\\
&\qquad\qquad\times
e^{
p\int_0^t
\gamma_1(u_0+W_s)\,dW_s
-
\frac12
\int_0^t
\bigl(
p^2\gamma_1^2(u_0+W_s)
+
2p\gamma_1(u_0+W_s)\gamma_2(u_0+W_s)
\bigr)\,ds
}
f_t(W)
\Bigg].
\end{align}

Let $\widehat W$ denote the process $W$ killed upon hitting
$\gamma-u_0$, and define a probability measure $\mathbb Q$ by
\[
\frac{d\mathbb Q}{d\mathbb P}
\Bigg|_{\mathcal F_t}
=
\exp
\left(
\int_0^t
\gamma_2(u_0+\widehat W_s)\,dW_s
-
\frac12
\int_0^t
\gamma_2^2(u_0+\widehat W_s)\,ds
\right).
\]

Indeed, evaluating \eqref{martp} at $p=0$ gives
\[
\mathbb E
\exp
\left(
\int_0^t
\gamma_2(u_0+\widehat W_s)\,dW_s
-
\frac12
\int_0^t
\gamma_2^2(u_0+\widehat W_s)\,ds
\right)
=1,
\]
showing that the above density has unit expectation.

Moreover,
\[
f_t(W)
=
\xi_t^p,
\qquad
\xi_t
=
\frac{
\mathbf 1_{\{t<\sigma^W_{\gamma-u_0}\}}
}{
\varphi^{-1}(u_0+W_t)
}.
\]

By Girsanov's theorem,
\[
\widetilde W_t
=
W_t
-
\int_0^t
\gamma_2(u_0+\widehat W_s)\,ds
\]
is a Brownian motion under $\mathbb Q$.
Consequently,
\begin{align*}
&
\mathbb E
\left[
\left(
\frac{S_t^{1-\beta}}{1-\beta}
e^{\frac{\alpha^2t}{2}+i\alpha V_t}
\right)^p
\right]
\\
&=
\mathbb E_{\mathbb Q}
\Bigg[
e^{
-\frac12
p^2
\int_0^t
\gamma_1^2(u_0+W_s)\,ds
}
\left(
e^{
\int_0^t
\gamma_1(u_0+W_s)\,dW_s
-
\int_0^t
\gamma_1(u_0+W_s)\gamma_2(u_0+W_s)\,ds
}
\xi_t
\right)^p
\Bigg].
\end{align*}

Since $\gamma_1(x)\le\alpha$ for
$x\in(-\infty,\gamma)$,
\[
-\frac12
\int_0^t
\gamma_1^2(u_0+W_s)\,ds
=
-\frac12
\alpha^2t
+
\frac12
\int_0^t
\bigl(
\alpha^2-\gamma_1^2(u_0+W_s)
\bigr)\,ds.
\]

Therefore,
\begin{align*}
&
\mathbb E
\left[
\left(
\frac{S_t^{1-\beta}}{1-\beta}
e^{\frac{\alpha^2t}{2}+i\alpha V_t}
\right)^p
\right]
\\
&=
\mathbb E_{\mathbb Q}
\Bigg[
\Bigg(
\xi_t
\exp
\Bigg(
i\alpha\widetilde V_t
+
V_{\int_0^t
(\alpha^2-\gamma_1^2(u_0+W_s))\,ds}
\\
&\qquad\qquad\qquad
+
\int_0^t
\gamma_1(u_0+W_s)\,dW_s
-
\int_0^t
\gamma_1(u_0+W_s)\gamma_2(u_0+W_s)\,ds
\Bigg)
\Bigg)^p
\Bigg],
\end{align*}
where $V$ and $\widetilde V$ are independent Brownian motions,
independent of $W$.

Finally, both sides are positive random variables with identical Mellin
transforms for every $p>0$. By uniqueness of the Mellin transform,
they coincide in distribution, proving
\eqref{rownosc wg rozkladu complex}.
\hfill$\square$

\begin{cor}\label{cor}
Let $\rho\in(-1,1)$. Then, for every $t>0$,
\begin{align*}
S_t^{1-\beta}
\overset{d}{=}
(1-\beta)
e^{-\frac{\alpha^2t}{2}}
\frac{1}{\varphi^{-1}(u_0+W_t)}
&
\exp\!\left(
V_{\int_0^t
(\alpha^2-\gamma_1^2(u_0+W_s))\,ds}
\right)
\\
&\times
\exp\!\left(
\int_0^t
\gamma_1(u_0+W_s)\,dW_s
-
\int_0^t
\gamma_1(u_0+W_s)\gamma_2(u_0+W_s)\,ds
\right)
\\
&\times
\mathbf 1_{\{t<\sigma^W_{\gamma-u_0}\}},
\end{align*}
where $W$ is a (possibly weak) solution of
\begin{equation}\label{WSDE}
\widetilde W_t
=
W_t
-
\int_0^t
\gamma_2(u_0+W_s)\,ds,
\qquad
t<\sigma^W_{\gamma-u_0},
\end{equation}
with $W_0=0$, $\widetilde W$ is a standard Brownian motion, and
$V$ is a Brownian motion independent of $W$.
Moreover, by the Engelbert--Schmidt existence theorem, a weak solution
to \eqref{WSDE} exists up to the explosion time
$\sigma^W_{\gamma-u_0}$.
\end{cor}
\noindent
{\bf Proof:} Taking absolute values on both sides of
\eqref{rownosc wg rozkladu complex} eliminates the factors
$e^{i\alpha V_t}$ and $e^{i\alpha\widetilde V_t}$, which have unit
modulus. This immediately yields the asserted distributional identity. Finally, the continuity of $\gamma_2$ implies, by the
Engelbert--Schmidt theory, the existence of a weak solution to
\eqref{WSDE} on the interval
$(0,\sigma^W_{\gamma-u_0})$.
\hfill $\square$

\ \\
We now complete the proof of Theorem~\ref{T22} by establishing the
representation \eqref{wzor 2 tw o rozkladzie} for the law of the process
$W$.

Recall from Corollary~\ref{cor} that $W$ is a (possibly weak) solution of
the SDE
\[
\widetilde W_t
=
W_t
-
\int_0^t
\gamma_2(u_0+W_s)\,ds,
\qquad
t<\sigma^W_{\gamma-u_0},
\]
where $\widetilde W$ is a standard Brownian motion.
The explosion time of this diffusion coincides with the first hitting time
of the boundary point $\gamma-u_0$, namely
$\sigma^W_{\gamma-u_0}$.

We now apply Theorem~3.2 of \cite{KaratzasRuf} with
\[
X=W,
\qquad
X^o=\widetilde W.
\]
Since $X^o$ is driftless, it is simply Brownian motion and therefore
cannot explode before reaching the boundary $\gamma-u_0$.
Consequently, for every Borel set
$\Delta\subset \mathcal C([0,t])$,
Theorem~3.2 in \cite{KaratzasRuf} yields
\begin{align*}
\mathbb P\!\left(
W_\cdot\in\Delta,\,
t<\sigma^W_{\gamma-u_0}
\right)
&=
\mathbb E\Bigg[
\exp\!\left(
\int_0^t
\gamma_2(u_0+\widetilde W_s)\,
d\widetilde W_s
-
\frac12
\int_0^t
\gamma_2^2(u_0+\widetilde W_s)\,
ds
\right)
\\
&\qquad\qquad\qquad\qquad\times
\mathbf 1_{\{
\widetilde W_\cdot\in\Delta,\,
t<\sigma^{\widetilde W}_{\gamma-u_0}
\}}
\Bigg].
\end{align*}
Since $\widetilde W$ is a standard Brownian motion, this is precisely
the representation \eqref{wzor 2 tw o rozkladzie}, completing the proof.
\hfill$\square$

\subsection{Proof of Theorem \ref{T222}}

We start from the representation established in Theorem~\ref{T1}:
\begin{align}\label{eq:start-proof-T222}
\mathbb{E}\!\left[\theta_t^{-p}\mathbf{1}_{\{t<\sigma\}}\right]
=
\mathbb{E}\!\left[
e^{\int_0^t\nu_p(\theta_s^o)\,dB_s^o
-\frac12\int_0^t\nu_p^2(\theta_s^o)\,ds}
(\theta_t^o)^{-p}
\mathbf{1}_{\{t<\sigma^o\}}
\right],
\end{align}
where
\[
\nu_p(x):=\frac{\mu_p(x)}{\sigma(x)},
\qquad x>0.
\]

The key observation is that $\nu_p$ depends affinely on $p$. More
precisely,
\[
\nu_p(x)
=
p\nu_1(x)+\nu_2(x),
\]
where
\[
\nu_1(x)
=
\frac{\alpha^2x-\alpha\rho x^2}{\sigma(x)},
\qquad
\nu_2(x)
=
\frac{1}{\sigma(x)}
\left(
\frac{2-\beta}{2(1-\beta)}x^3
-\alpha\rho x^2
\right).
\]

Consequently,
\begin{align*}
&
\int_0^t
\nu_p(\theta_s^o)\,dB_s^o
-
\frac12
\int_0^t
\nu_p^2(\theta_s^o)\,ds
\\
&=
p
\int_0^t
\nu_1(\theta_s^o)\,dB_s^o
+
\int_0^t
\nu_2(\theta_s^o)\,dB_s^o
-
\frac{p^2}{2}
\int_0^t
\nu_1^2(\theta_s^o)\,ds
\\
&\qquad
-
p
\int_0^t
\nu_1(\theta_s^o)\nu_2(\theta_s^o)\,ds
-
\frac12
\int_0^t
\nu_2^2(\theta_s^o)\,ds.
\end{align*}

Moreover, a direct computation shows that
\[
\nu_1(x)\leq \alpha,
\qquad x>0,
\]
so that
\[
-\frac{p^2}{2}
\int_0^t
\nu_1^2(\theta_s^o)\,ds
=
-\frac{\alpha^2p^2}{2}t
+
\frac{p^2}{2}
\int_0^t
\bigl(
\alpha^2-\nu_1^2(\theta_s^o)
\bigr)\,ds.
\]

Applying the generalized Girsanov theorem for explosive diffusions once
again, we introduce the probability measure
\[
\frac{d\mathbb Q}{d\mathbb P}\Big|_{\mathcal F_t}
=
\exp\!\left(
\int_0^t
\nu_2(\theta_s^o)\,dB_s^o
-
\frac12
\int_0^t
\nu_2^2(\theta_s^o)\,ds
\right)
\mathbf 1_{\{t<\sigma^o\}},
\]
under which
\[
W_t^o
=
B_t^o
-
\int_0^t
\nu_2(\theta_s^o)\,ds,
\qquad
t<\sigma^o,
\]
is a Brownian motion.

Furthermore, under $\mathbb Q$ the process $\theta^o$ satisfies
\[
d\theta_t^o
=
\sigma(\theta_t^o)\,dW_t^o
+
\sigma(\theta_t^o)\nu_2(\theta_t^o)\,dt,
\qquad
t<\sigma^o.
\]
Since the coefficients are continuous, the existence of a weak solution
follows from the classical Engelbert--Schmidt theory.

Substituting the above decomposition into
\eqref{eq:start-proof-T222} yields
\begin{align*}
&
\mathbb E
\left[
\theta_t^{-p}
\mathbf 1_{\{t<\sigma\}}
\right]
\\
&=
e^{-\frac{\alpha^2p^2}{2}t}
\,
\mathbb E_{\mathbb Q}
\Bigg[
\exp\!\left(
p
\int_0^t
\nu_1(\theta_s^o)\,dW_s^o
+
\frac{p^2}{2}
\int_0^t
(\alpha^2-\nu_1^2(\theta_s^o))\,ds
\right)
\\
&\hspace{3.8cm}\times
\exp\!\left(
-
p
\int_0^t
\nu_1(\theta_s^o)\nu_2(\theta_s^o)\,ds
\right)
(\theta_t^o)^{-p}
\mathbf 1_{\{t<\sigma^o\}}
\Bigg].
\end{align*}

Finally, using the elementary identity
\[
\mathbb E e^{pV_A}
=
e^{\frac12p^2A},
\]
valid for every Brownian motion $V$ independent of
$(W^o,\theta^o)$ and every non-negative random variable
$A$ measurable with respect to $(W^o,\theta^o)$, we obtain
\begin{align*}
&
\mathbb E
\left[
\theta_t^{-p}
\mathbf 1_{\{t<\sigma\}}
\right]
\\
&=
e^{-\frac{\alpha^2p^2}{2}t}
\,
\mathbb E_{\mathbb Q}
\Bigg[
\exp\!\left(
p
\int_0^t
\nu_1(\theta_s^o)\,dW_s^o
+
p
V_{\int_0^t
(\alpha^2-\nu_1^2(\theta_s^o))\,ds}
\right)
\\
&\hspace{3.8cm}\times
\exp\!\left(
-
p
\int_0^t
\nu_1(\theta_s^o)\nu_2(\theta_s^o)\,ds
\right)
(\theta_t^o)^{-p}
\mathbf 1_{\{t<\sigma^o\}}
\Bigg].
\end{align*}

The desired distributional identity now follows from the uniqueness
of the Mellin transform exactly as in the proof of
Theorem~\ref{T22}:
\[
S_t^{1-\beta}
\overset{d}{=}
(1-\beta)
e^{-\frac{\alpha^2t}{2}}
e^{\int_0^t
\nu_1(\theta_s^o)\,dW_s^o
+
V_{\int_0^t
(\alpha^2-\nu_1^2(\theta_s^o))\,ds}}
(\theta_t^o)^{-1}
\mathbf 1_{\{t<\sigma^o\}},
\]
which completes the proof.
\hfill$\square$

\subsection{Proof of Corollary \ref{cor:mixture-bs}}

By Theorem~\ref{T222},
\[
    S_t^q
    \overset{d}{=}
    \exp\{A_t+V_{\Sigma_t^2}\}
    \mathbf 1_{\{t<\sigma^o\}}.
\]
Conditionally on the path of $(\theta^o,W^o)$, the random variable
$V_{\Sigma_t^2}$ has the same law as $\Sigma_t Z$, where $Z$ is a
standard Gaussian random variable independent of $(\theta^o,W^o)$. Hence,
on $\{t<\sigma^o\}$,
\[
    S_t
    \overset{d}{=}
    \exp\left(\frac{A_t}{q}+\frac{\Sigma_t}{q}Z\right).
\]
Set
\[
    \mathcal V_t:=\frac{\Sigma_t}{q}
\]
and choose $\mathcal S_t$ such that
\[
    \log \mathcal S_t-\frac12\mathcal V_t^2
    =
    \frac{A_t}{q}.
\]
This gives
\[
    \mathcal S_t
    =
    \exp\left(
        \frac{A_t}{q}
        +
        \frac{\Sigma_t^2}{2q^2}
    \right).
\]
Therefore, conditionally on $(\theta^o,W^o)$,
\[
    S_t
    \overset{d}{=}
    \mathcal S_t
    \exp\left(
        -\frac12\mathcal V_t^2+\mathcal V_t Z
    \right).
\]
The conditional call payoff is thus exactly the Black--Scholes formula
with spot $\mathcal S_t$, strike $K$, zero interest rate, and total
volatility $\mathcal V_t$:
\[
    \mathbb E\left[(S_t-K)^+\mid \theta^o,W^o\right]
    =
    BS(\mathcal S_t,K,\mathcal V_t).
\]
Multiplying by $\mathbf 1_{\{t<\sigma^o\}}$ and taking expectation gives
the result. \hfill $\square$


\bibliographystyle{amsplain}
\nocite{*}
\bibliography{sabr}

\end{document}